\documentclass[aps,prl,superscriptaddress,showpacs,floatfix,twocolumn]{revtex4-2}

\usepackage{graphicx}   

\begin{document}

\title{Single Electrons from Heavy Flavor Decays in p+p Collisions
at $\sqrt{s} = 200$~GeV}

\newcommand{\abilene}{Abilene Christian University, Abilene, TX 79699, USA}
\newcommand{\acadsin}{Institute of Physics, Academia Sinica, Taipei 11529, Taiwan}
\newcommand{\banaras}{Department of Physics, Banaras Hindu University, Varanasi 221005, India}
\newcommand{\barc}{Bhabha Atomic Research Centre, Bombay 400 085, India}
\newcommand{\bnl}{Brookhaven National Laboratory, Upton, NY 11973-5000, USA}
\newcommand{\caucr}{University of California - Riverside, Riverside, CA 92521, USA}
\newcommand{\ciae}{China Institute of Atomic Energy (CIAE), Beijing, People's Republic of China}
\newcommand{\cns}{Center for Nuclear Study, Graduate School of Science, University of Tokyo, 7-3-1 Hongo, Bunkyo, Tokyo 113-0033, Japan}
\newcommand{\columbia}{Columbia University, New York, NY 10027 and Nevis Laboratories, Irvington, NY 10533, USA}
\newcommand{\dapnia}{Dapnia, CEA Saclay, F-91191, Gif-sur-Yvette, France}
\newcommand{\debrecen}{Debrecen University, H-4010 Debrecen, Egyetem t{\'e}r 1, Hungary}
\newcommand{\fsu}{Florida State University, Tallahassee, FL 32306, USA}
\newcommand{\gsu}{Georgia State University, Atlanta, GA 30303, USA}
\newcommand{\hiroshima}{Hiroshima University, Kagamiyama, Higashi-Hiroshima 739-8526, Japan}
\newcommand{\ihepprot}{IHEP Protvino, State Research Center of Russian Federation, Institute for High Energy Physics, Protvino, 142281, Russia}
\newcommand{\isu}{Iowa State University, Ames, IA 50011, USA}
\newcommand{\jinrdubna}{Joint Institute for Nuclear Research, 141980 Dubna, Moscow Region, Russia}
\newcommand{\kaeri}{KAERI, Cyclotron Application Laboratory, Seoul, South Korea}
\newcommand{\kangnung}{Kangnung National University, Kangnung 210-702, South Korea}
\newcommand{\kek}{KEK, High Energy Accelerator Research Organization, Tsukuba, Ibaraki 305-0801, Japan}
\newcommand{\kfki}{KFKI Research Institute for Particle and Nuclear Physics of the Hungarian Academy of Sciences (MTA KFKI RMKI), H-1525 Budapest 114, POBox 49, Budapest, Hungary}
\newcommand{\korea}{Korea University, Seoul, 136-701, Korea}
\newcommand{\kurchatov}{Russian Research Center ``Kurchatov Institute", Moscow, Russia}
\newcommand{\kyoto}{Kyoto University, Kyoto 606-8502, Japan}
\newcommand{\labllr}{Laboratoire Leprince-Ringuet, Ecole Polytechnique, CNRS-IN2P3, Route de Saclay, F-91128, Palaiseau, France}
\newcommand{\lawllnl}{Lawrence Livermore National Laboratory, Livermore, CA 94550, USA}
\newcommand{\losalamos}{Los Alamos National Laboratory, Los Alamos, NM 87545, USA}
\newcommand{\lpc}{LPC, Universit{\'e} Blaise Pascal, CNRS-IN2P3, Clermont-Fd, 63177 Aubiere Cedex, France}
\newcommand{\lund}{Department of Physics, Lund University, Box 118, SE-221 00 Lund, Sweden}
\newcommand{\muenster}{Institut f\"ur Kernphysik, University of Muenster, D-48149 Muenster, Germany}
\newcommand{\myongji}{Myongji University, Yongin, Kyonggido 449-728, Korea}
\newcommand{\nagasaki}{Nagasaki Institute of Applied Science, Nagasaki-shi, Nagasaki 851-0193, Japan}
\newcommand{\newmex}{University of New Mexico, Albuquerque, NM 87131, USA}
\newcommand{\nmsu}{New Mexico State University, Las Cruces, NM 88003, USA}
\newcommand{\ornl}{Oak Ridge National Laboratory, Oak Ridge, TN 37831, USA}
\newcommand{\orsay}{IPN-Orsay, Universite Paris Sud, CNRS-IN2P3, BP1, F-91406, Orsay, France}
\newcommand{\pnpi}{PNPI, Petersburg Nuclear Physics Institute, Gatchina,  Leningrad region, 188300, Russia}
\newcommand{\riken}{RIKEN, The Institute of Physical and Chemical Research, Wako, Saitama 351-0198, Japan}
\newcommand{\rikjrbrc}{RIKEN BNL Research Center, Brookhaven National Laboratory, Upton, NY 11973-5000, USA}
\newcommand{\saispbstu}{Saint Petersburg State Polytechnic University, St. Petersburg, Russia}
\newcommand{\saopaulo}{Universidade de S{\~a}o Paulo, Instituto de F\'{\i}sica, Caixa Postal 66318, S{\~a}o Paulo CEP05315-970, Brazil}
\newcommand{\seoulnat}{System Electronics Laboratory, Seoul National University, Seoul, South Korea}
\newcommand{\stonybrkc}{Chemistry Department, Stony Brook University, SUNY, Stony Brook, NY 11794-3400, USA}
\newcommand{\stonycrkp}{Department of Physics and Astronomy, Stony Brook University, SUNY, Stony Brook, NY 11794, USA}
\newcommand{\subatech}{SUBATECH (Ecole des Mines de Nantes, CNRS-IN2P3, Universit{\'e} de Nantes) BP 20722 - 44307, Nantes, France}
\newcommand{\tenn}{University of Tennessee, Knoxville, TN 37996, USA}
\newcommand{\titech}{Department of Physics, Tokyo Institute of Technology, Tokyo, 152-8551, Japan}
\newcommand{\tsukuba}{Institute of Physics, University of Tsukuba, Tsukuba, Ibaraki 305, Japan}
\newcommand{\vandy}{Vanderbilt University, Nashville, TN 37235, USA}
\newcommand{\waseda}{Waseda University, Advanced Research Institute for Science and Engineering, 17 Kikui-cho, Shinjuku-ku, Tokyo 162-0044, Japan}
\newcommand{\weizmann}{Weizmann Institute, Rehovot 76100, Israel}
\newcommand{\yonsei}{Yonsei University, IPAP, Seoul 120-749, Korea}
\affiliation{\abilene}
\affiliation{\acadsin}
\affiliation{\banaras}
\affiliation{\barc}
\affiliation{\bnl}
\affiliation{\caucr}
\affiliation{\ciae}
\affiliation{\cns}
\affiliation{\columbia}
\affiliation{\dapnia}
\affiliation{\debrecen}
\affiliation{\fsu}
\affiliation{\gsu}
\affiliation{\hiroshima}
\affiliation{\ihepprot}
\affiliation{\isu}
\affiliation{\jinrdubna}
\affiliation{\kaeri}
\affiliation{\kangnung}
\affiliation{\kek}
\affiliation{\kfki}
\affiliation{\korea}
\affiliation{\kurchatov}
\affiliation{\kyoto}
\affiliation{\labllr}
\affiliation{\lawllnl}
\affiliation{\losalamos}
\affiliation{\lpc}
\affiliation{\lund}
\affiliation{\muenster}
\affiliation{\myongji}
\affiliation{\nagasaki}
\affiliation{\newmex}
\affiliation{\nmsu}
\affiliation{\ornl}
\affiliation{\orsay}
\affiliation{\pnpi}
\affiliation{\riken}
\affiliation{\rikjrbrc}
\affiliation{\saispbstu}
\affiliation{\saopaulo}
\affiliation{\seoulnat}
\affiliation{\stonybrkc}
\affiliation{\stonycrkp}
\affiliation{\subatech}
\affiliation{\tenn}
\affiliation{\titech}
\affiliation{\tsukuba}
\affiliation{\vandy}
\affiliation{\waseda}
\affiliation{\weizmann}
\affiliation{\yonsei}
\author{S.S.~Adler} \affiliation{\bnl}
\author{S.~Afanasiev}   \affiliation{\jinrdubna}
\author{C.~Aidala}  \affiliation{\bnl}
\author{N.N.~Ajitanand} \affiliation{\stonybrkc}
\author{Y.~Akiba}   \affiliation{\kek} \affiliation{\riken}
\author{J.~Alexander}   \affiliation{\stonybrkc}
\author{R.~Amirikas}    \affiliation{\fsu}
\author{L.~Aphecetche}  \affiliation{\subatech}
\author{S.H.~Aronson}   \affiliation{\bnl}
\author{R.~Averbeck}    \affiliation{\stonycrkp}
% ...
\collaboration{PHENIX Collaboration} \noaffiliation

\date{\today}

\begin{abstract}

The invariant differential cross section for inclusive electron
production in $p + p$ collisions at $\sqrt{s} = 200$~GeV has been
measured by the PHENIX experiment at the Relativistic Heavy Ion
Collider over the transverse momentum range 
$0.4 \le p_T \le 5.0$~GeV/$c$ in the central rapidity region 
($|\eta| \le 0.35$).  The contribution to the inclusive electron 
spectrum from semileptonic decays of hadrons carrying heavy flavor, 
{\it i.e.} charm quarks or, at high $p_T$, bottom quarks, is determined via 
three independent methods.  The resulting electron spectrum from heavy 
flavor decays is compared to recent leading and next-to-leading order 
perturbative QCD calculations.  The total cross section of charm 
quark-antiquark pair production is determined to be 
$\sigma_{c\bar{c}} = 0.92 \pm 0.15 {\rm (stat.)} \pm 0.54 {\rm (sys.)}$~mb. 

\end{abstract}


\maketitle

The production of hadrons carrying heavy quarks, {\it i.e.} charm or bottom,
serves as a crucial proving ground for quantum chromodynamics (QCD), the
theory of the strong interaction.
Because of the large quark masses, charm and bottom production can be treated
by perturbative QCD (pQCD) even at small momenta without being significantly
affected by additional soft processes~\cite{mangano93}.
This is in distinct contrast to the production of particles composed solely of
light quarks, which can be evaluated perturbatively only for sufficiently
large momenta.
Consequently, pQCD calculations of heavy quark production are expected to be
reliable over the full momentum range experimentally accessible at collider
energies.

For bottom production, next-to-leading order (NLO) calculations are
in reasonable agreement with data~\cite{cacciari04mangano04}.  Charm
measurements at $\sqrt{s} = 1.96$~TeV exist for high transverse
momentum ($p_T$) only~\cite{acosta03}, where the cross section is
higher than NLO predictions by $\ge 50$\%.  However, these
discrepancies are within the substantial experimental and
theoretical uncertainties~\cite{acosta03}.  At the Relativistic Heavy
Ion Collider (RHIC), charm yields have been measured for $p + p$
and $d + Au$ collisions at $\sqrt{s_{NN}} =
200$~GeV~\cite{star_dau,phenix_dau} as well as for $Au + Au$
collisions at 130 and 200~GeV~\cite{phenix_auau130,phenix_auau200}.
Further measurements are crucial for a better understanding of heavy
flavor production at RHIC.  In particular, the relevance of higher
order processes and other production mechanisms like jet
fragmentation is unclear.

We report on the central rapidity production ($|\eta| \le 0.35$) of
inclusive electrons, $(e^+ + e^-)/2$, in $p + p$ collisions at
$\sqrt{s} = 200$~GeV measured by the PHENIX
experiment~\cite{phenix_nim} at RHIC.  Contributions from
semileptonic heavy flavor decays are extracted in the electron $p_T$
range $0.4 \le p_T \le 5.0$~GeV/$c$.  The resulting invariant
differential cross section is an important benchmark for pQCD
calculations of heavy quark production.  Furthermore, it provides a
crucial baseline for measurements in nuclear collisions at RHIC.
Since hadronic heavy flavor production is expected to be dominated
by initial parton scattering, systematic studies in $p + p$ and $d +
Au$ collisions should be sensitive to the nucleon parton
distribution functions as well as to their nuclear modifications
such as shadowing~\cite{lin96}.  In $Au + Au$ collisions, heavy
quarks constitute a unique and, with the data presented here,
calibrated probe for the hot and dense medium created in the
collisions.  Possible medium modification of heavy flavor probes
include energy loss~\cite{dokshitzer01,armesto05}, azimuthal
asymmetry~\cite{lin03greco04}, and quarkonia
suppression~\cite{matsui86} or enhancement~\cite{pbm,thews}.

The data used here were recorded by PHENIX during RHIC Run-2.
Beam-beam counters (BBC), positioned at pseudorapidities
$3.1 < |\eta| < 3.9$, measured the collision vertex and provided the
minimum bias (MB) interaction trigger defined by at least one hit on each side
of the vertex.
Events containing high $p_T$ electrons were selected by an additional level-1
trigger in coincidence with the MB trigger.
This level-1 trigger required a minimum energy deposit of 0.75~GeV in a
$2 \times 2$ tile of towers in the electromagnetic calorimeter
(EMC)~\cite{phenix_pi0pp}.
After a vertex cut of $|z_{vtx}| < 20$~cm, an equivalent of
$465 \times 10^6$ MB events sampled by the EMC trigger was analyzed in
addition to the $15 \times 10^6$ events recorded with the MB trigger itself.

The PHENIX east arm spectrometer ($|\eta| < 0.35$, $\Delta \phi =
\pi/2$) includes a drift chamber and a pad chamber layer for charged
particle tracking.  Tracks were confirmed by hits in the EMC matching
in position with the track projection within $3\sigma$.  Electron
candidates required at least two associated hits in the ring imaging
\v{C}erenkov detector (RICH) in the projected ring area.  Random
coincidences of hadron tracks and hits in the RICH occurred with a
probability of $(3.0\pm1.5) \times 10^{-4}$.  For electrons the
energy $E$ deposited in the EMC must be consistent with the momentum
$p$.  Requiring $|(E-p)/p| < 3\sigma$, a total charged hadron
rejection factor of about 10$^4$ (10$^5$) was achieved for $p_T =
0.4$ $(\ge 2.0)$~GeV/$c$.  Remaining background ($<1$~\%) was
measured via event mixing and subtracted statistically.

The differential cross section for electron production was calculated as
\begin{equation}
E \frac{d^3\sigma}{dp^3} = \frac{1}{\epsilon_{bias} \int{{\cal{L}}dt}}
               \frac{N_e}{2\pi p_T \Delta y \Delta p_T}
               \frac{1}{A \epsilon_{rec}},
\end{equation}
where $\int{{\cal{L}}dt}$ is the integrated luminosity measured with the MB
trigger or sampled with the EMC trigger, respectively, $\epsilon_{bias}$
is the probability for an electron event to fulfill the MB trigger condition,
$N_e$ is the measured electron yield, and $A \epsilon_{rec}$ is the product of
geometrical acceptance and reconstruction efficiency.
For the EMC triggered sample, $\epsilon_{rec}$ includes the trigger efficiency
$\epsilon_{lvl1}$.

$\int{{\cal{L}}dt}$ is calculated as $N_{MB}/\sigma_{BBC}$, where $N_{MB}$
is the number of MB triggers or, for the EMC triggered sample, the number
of EMC triggers divided by the measured fraction of MB events which
simultaneously fulfill the EMC trigger criterion.
With the MB trigger cross section
$\sigma_{BBC} = 21.8 \pm 2.1$~mb~\cite{phenix_pi0pp}, the analyzed data
samples correspond to integrated luminosities of 0.7 nb$^{-1}$ (MB trigger)
and 21 nb$^{-1}$ (EMC trigger), respectively.
The $p_T$ independent trigger bias $\epsilon_{bias} = 0.75 \pm 0.02$ was
measured for events containing a $\pi^0$ with
$p_T > 1.5$~GeV/$c$~\cite{phenix_pi0pp} and confirmed for charged hadrons with
$p_T > 0.2$~GeV/$c$~\cite{phenix_ppg050}, indicating a universal bias both for
hard and soft processes.
$A \epsilon_{rec}$ was calculated as a function of $p_T$ ($< 10$~\% variation
over the full $p_T$ range) in a GEANT~\cite{geant} simulation of electrons
with flat distributions in rapidity ($|y| < 0.6$), azimuth ($0 < \phi < 2\pi$),
and event vertex ($|z| < 30$~cm) as input.
The simulated detector response was carefully tuned to match the real detector.
Rigorous fiducial cuts were applied to eliminate active area mismatches
between data and simulation as well as run-by-run variations.
The trigger efficiency $\epsilon_{lvl1}$, evaluated for single electrons
in the fiducial area, rises from zero at low $p_T$ to $95 \pm 5$\% for
$p_T > 2$~GeV/$c$.
Finally the effect of finite bin width in $p_T$ was appropriately corrected
for.

\begin{figure}[t]
\includegraphics[width=1.0\linewidth]{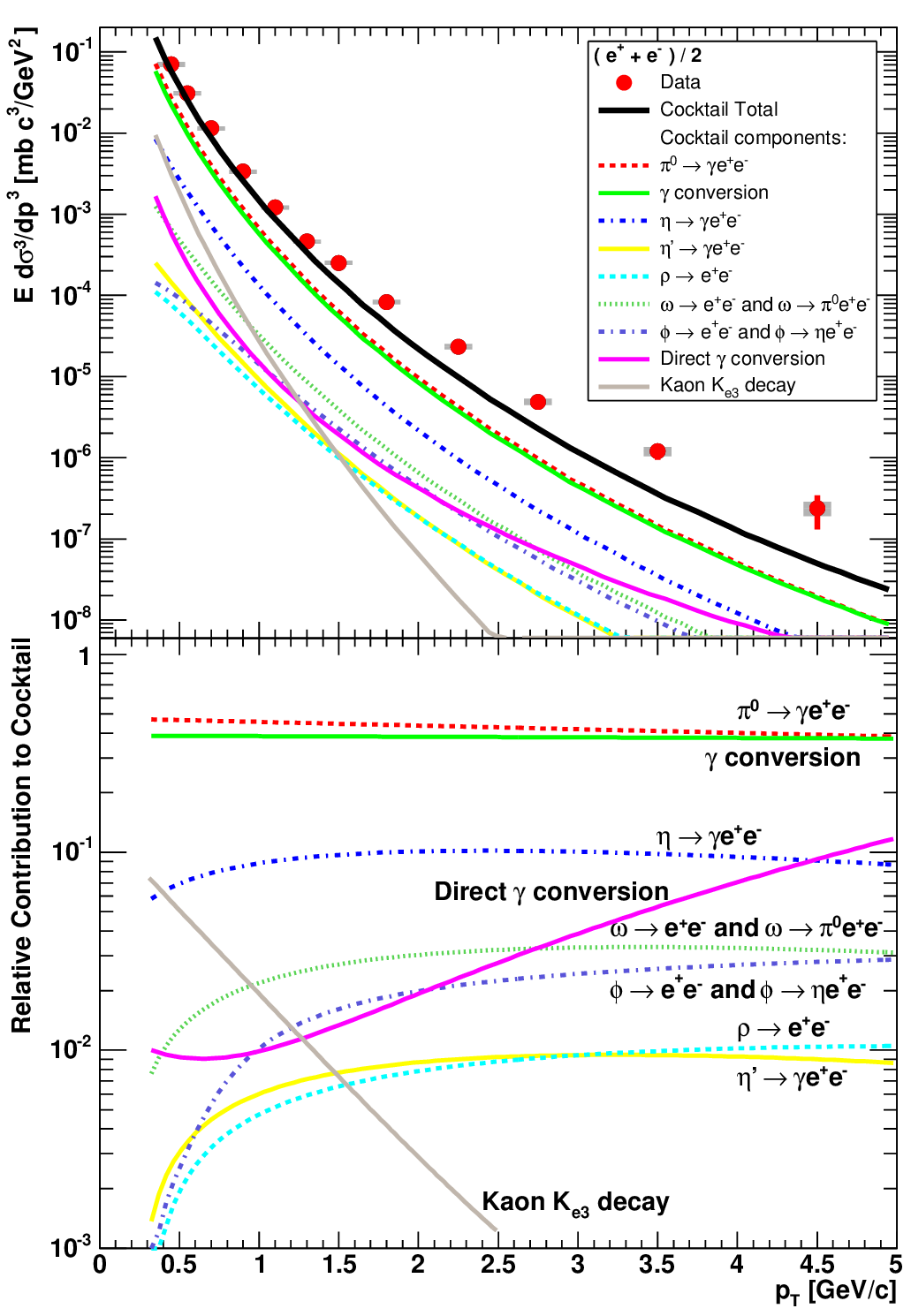} 
\caption{\label{fig:PPG037_Fig1}
(Color online) (a) Inclusive electron invariant differential cross
section, measured in $p + p$ collisions at $\sqrt{s} = 200$~GeV,
compared with all contributions from electron sources included in
the background {\it cocktail}.  Error bars (boxes) correspond to
statistical (systematic) uncertainties.  (b) Relative contributions
of all electron sources to the background {\it cocktail}.}
\end{figure}

The corrected electron spectra from the MB and EMC triggered samples
cover $p_T$ ranges of $0.4 < p_T < 2.0$~GeV/$c$ and $0.6 < p_T <
5.0$~GeV/$c$, respectively.  They are consistent with each other
within the statistical uncertainties in the $p_T$ region of overlap.
The weighted average of both measurements is shown in
Fig.~\ref{fig:PPG037_Fig1}(a).

The systematic uncertainty of the inclusive electron spectrum is about 12\%,
almost $p_T$ independent, calculated as the sum in quadrature of contributions
from the acceptance calculation (7\%), electron identification cuts (5.2\%),
run-by-run variations (4\%), tracking efficiency (3\%), momentum scale
(1 - 5\%), and other smaller uncertainties (more details on the uncertainty
estimations can be found in~\cite{sb_thesis}).  The value of 12\% does not 
include the 9.6\% uncertainty of the absolute normalization.

The invariant cross section of electrons from heavy flavor decays
was determined by subtracting a {\it cocktail} of contributions from
other sources from the inclusive data.  The most important background
is the $\pi^0$ Dalitz decay which was calculated with a hadron decay
generator using a parameterization of measured
$\pi^0$~\cite{phenix_pi0pp} and $\pi^\pm$~\cite{phenix_pichargepp}
spectra as input.  The spectral shapes of other light hadrons $h$
were obtained from the pion spectra by $m_T$ scaling.  Within this
approach the ratios $h/\pi^0$ are constant at high $p_T$ and for the
relative normalization we used: $\eta/\pi^0 = 0.45 \pm
0.10$~\cite{phenix_ppg051}, $\rho/\pi^0 = 1.0 \pm 0.3$,
$\omega/\pi^0 = 1.0 \pm 0.3$, $\eta'/\pi^0 = 0.25 \pm 0.08$, and
$\phi/\pi^0 = 0.40 \pm 0.12$.  Only the $\eta$ contribution is of any
practical relevance.  Another major electron source is the conversion
of photons, mainly from $\pi^0 \rightarrow \gamma\gamma$ decays, in
material within the acceptance.  The spectra of electrons from
conversions and Dalitz decays are very similar.  In a GEANT
simulation of $\pi^0$ decays, the ratio of electrons from
conversions to electrons from Dalitz decays was determined to be
$0.73 \pm 0.07$, essentially $p_T$ independent.  Contributions from
photon conversions from other sources were taken into account as
well.  In addition, electrons from kaon decays ($K_{e3}$), determined
in a GEANT simulation based on measured kaon
spectra~\cite{phenix_pichargepp}, and electrons from external as
well as internal conversions of direct
photons~\cite{vogelsang,phenix_directgammapp} were considered in the
cocktail.  All background sources are compared with the inclusive
data in Fig.~\ref{fig:PPG037_Fig1}(a) with the relative
contributions shown in Fig.~\ref{fig:PPG037_Fig1}(b).  The total
systematic uncertainty of the cocktail is about 12\%, essentially
$p_T$ independent.  This uncertainty is dominated by the systematic
error of the pion parameterization ($\approx 10$\%).  Other
systematic uncertainties, mainly the $\eta/\pi^0$ normalization and,
at high $p_T$, the contribution from direct radiation, are much
smaller.

\begin{figure}[t]
\includegraphics[width=1.0\linewidth,clip]{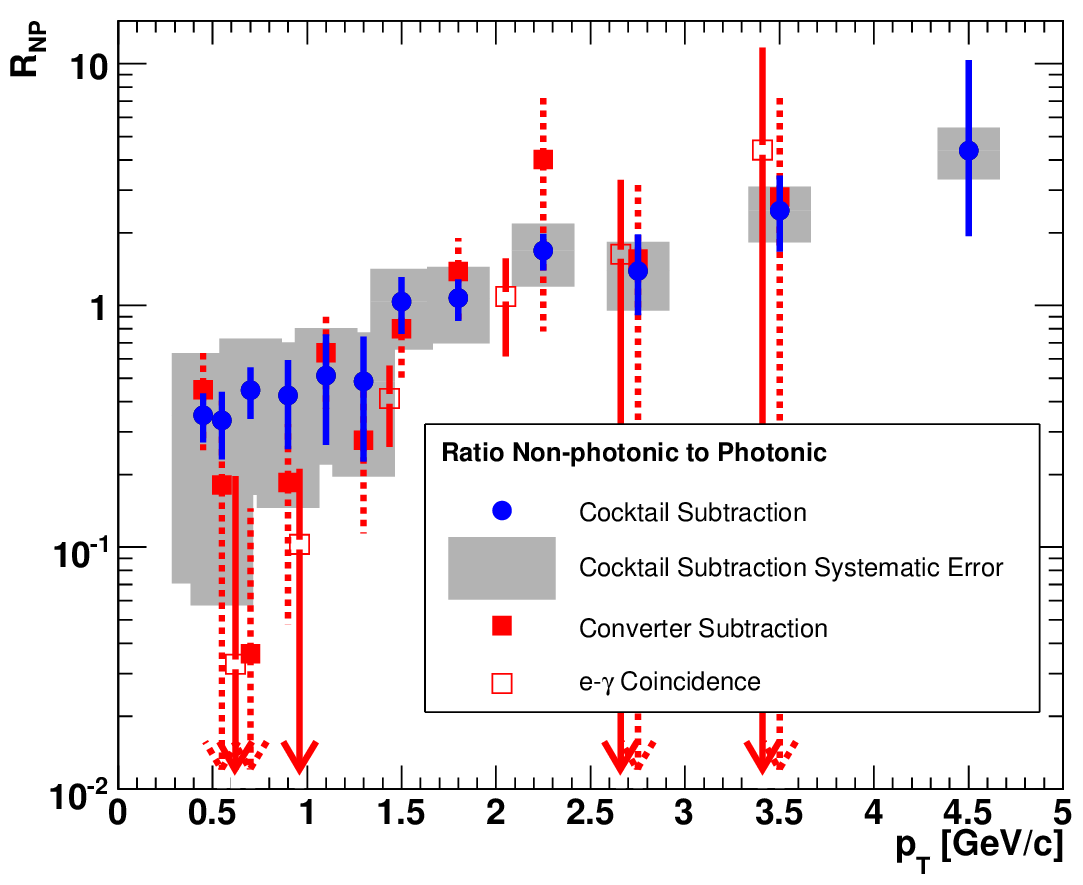} \caption {\label{fig:PPG037_Fig2}
Ratio of electrons from heavy flavor decays (non-photonic) and other
sources (photonic), $R_{NP}$, for three independent analysis
methods.  Error bars (boxes) are statistical ({\it cocktail}
systematic) uncertainties.}
\end{figure}

Given the small amount of material within the acceptance (Be beam
pipe: 0.29~\% $X_0$; air: 0.28~\% $X_0$) the ratio $R_{NP}$ of
non-photonic electrons from heavy flavor decays to background from
photonic sources is large ($R_{NP} > 1$ for $p_T > 1.5$~GeV/$c$) as
shown in Fig.~\ref{fig:PPG037_Fig2}.  Two complementary analysis
methods confirm the {\it cocktail} result:

The {\it converter} technique~\cite{phenix_auau200} compares
electron spectra measured with an additional photon converter
$X_{C} = 1.67$~\% $X_0$ introduced into the acceptance to measurements
without converter.
The converter increases the contribution from conversions and Dalitz
decays by a fixed factor, which was determined precisely via GEANT
simulations.
Thus, the electron spectra from photonic and non-photonic sources can be
deduced (Fig.~\ref{fig:PPG037_Fig2}).
The drawbacks of the {\it converter} method are the limitation in statistics
of the converter run period and the fact that the photonic contribution is
small at high $p_T$.

The {\it $e\gamma$ coincidence} technique evaluates the correlation
of electrons and photons via their invariant mass.  Electrons from
$\pi^0$ Dalitz decays or the conversion of one of the photons from
$\pi^0 \rightarrow \gamma\gamma$ decays are correlated with a
photon, in contrast to electrons from semileptonic heavy flavor
decays.  Comparing the measured $e\gamma$ coincidence rate with the
simulated rate for single $\pi^0$ events, allows $R_{NP}$ to be
deduced as shown in Fig.~\ref{fig:PPG037_Fig2}, once corrections for
contributions from other photonic sources are applied.

\begin{figure}[t]
\includegraphics[width=1.0\linewidth]{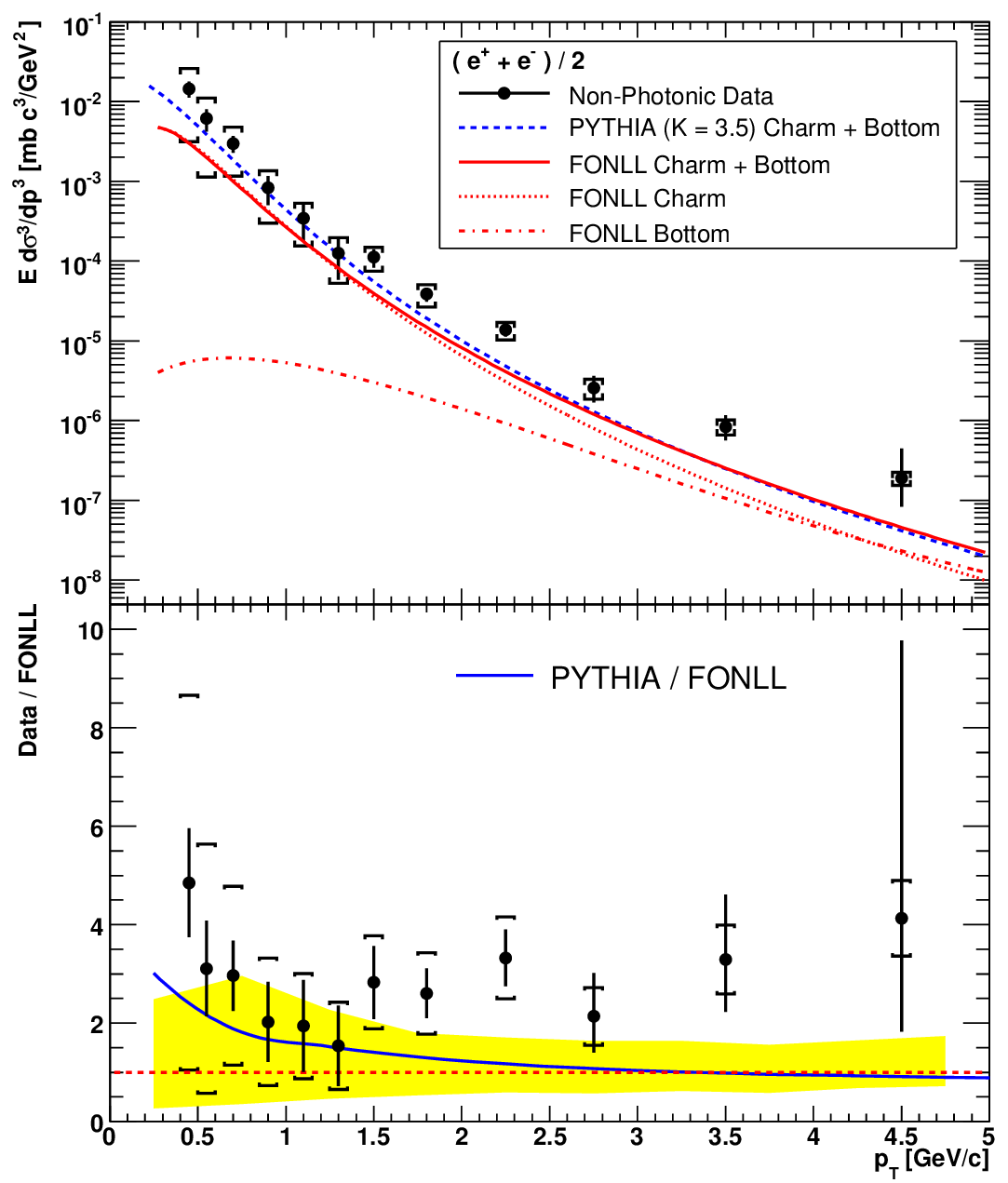}
\caption {\label{fig:PPG037_Fig3} (a) Invariant differential cross
section of electrons from heavy flavor decays compared with PYTHIA
LO (with $K = 3.5$) and FONLL pQCD calculations.  Error bars
(brackets) show statistical (systematic) uncertainties.  For the
FONLL calculation contributions from charm, bottom and bottom
cascade decays are shown separately.  (b) Ratio of data and FONLL
calculation with experimental statistical (error bars) and
systematic (brackets) uncertainties as well as the theoretical
uncertainty (grey band).  The solid line corresponds to the ratio of
PYTHIA and FONLL.}
\end{figure}

After subtracting the background cocktail from the inclusive
electron spectrum the invariant differential cross section of
electrons from heavy flavor decays is shown in
Fig.~\ref{fig:PPG037_Fig3} compared with two theoretical
predictions.  A leading order (LO) PYTHIA calculation, tuned to
existing charm and bottom hadroproduction
measurements~\cite{pythia_tuned}, is in reasonable agreement with
the data for $p_T < 1.5$~GeV/$c$, but underestimates the cross
section at higher $p_T$.  It is important to note that this
calculation includes a scale factor $K = 3.5$ to accomodate for
neglected NLO contributions.  A {\it Fixed-Order plus
Next-to-Leading-Log} (FONLL) pQCD calculation~\cite{cacciari05}
still leaves room for further contributions beyond the included NLO
processes.  The predicted contribution from bottom decays is
irrelevant for the electron cross section at $p_T < 3$~GeV/$c$ and
becomes significant only for $p_T > 4$~GeV/$c$.

 The charm production cross section was derived from the integrated
electron cross section for $p_{T} > p_{T,low} = 0.6 (0.8)$~GeV/$c$
($d\sigma_{e}^{p_{T,low}}/dy = 4.78 (2.15) \pm 0.78 (0.46) {\rm
(stat.)} \pm 1.74 (0.68) {\rm (sys.)}$~$\times 10^{-3}$~mb).  Since
in the low $p_T$ region, which dominates the total cross section,
PYTHIA describes the measured spectrum reasonably well, the total
charm cross section was determined by extrapolating the properly
scaled PYTHIA spectrum to $p_T = 0$~GeV/$c$.  First the PYTHIA
spectra for electrons from charm and bottom decays were fitted to
the data for $p_T > 0.6$~GeV/$c$, with only the normalizations as
free parameters.  The resulting central rapidity charm production
cross section was determined to be $d\sigma_{c\bar{c}}/dy = 0.20 \pm
0.03 {\rm (stat.)} \pm 0.11 {\rm (sys.)}$~mb, where the systematic
error is dominated by the uncertainty of the electron spectrum
itself ($\approx$56\%), evaluated by refitting PYTHIA to the data at
the minimum and maximum of the 1$\sigma$ systematic error band.
Additional uncertainties from the relative ratios of different
charmed hadron species and their branching ratios into electrons
($\approx$9\%) and the variation of the PYTHIA spectral shape
($\approx$11\%)~\cite{phenix_auau200} were added in quadrature.  The
rapidity integrated cross section was determined to be
$\sigma_{c\bar{c}} = 0.92 \pm 0.15 {\rm (stat.)} \pm 0.54 {\rm
(sys.)}$~mb, where various parton distribution functions (GRV98LO
and MRST(c-g)~\cite{pdf} in addition to the default
CTEQ5L~\cite{cteq5l}) were used for the extrapolation, with an
associated extra systematic error of
$\approx$6\%~\cite{phenix_auau200} added in quadrature.

Within errors the integrated charm cross section is compatible with
data from $Au + Au$ collisions~\cite{phenix_auau200} (minimum bias
value: $0.622 \pm 0.057 \pm 0.160$~mb per $NN$ collision) and from
$d + Au$ collisions~\cite{star_dau} ($1.3 \pm 0.2 \pm 0.4$~mb) at
the same $\sqrt{s_{NN}} = 200$~GeV.  The FONLL cross section is
smaller ($\sigma_{c\bar{c}}^{FONLL} = 0.256^{+0.400}_{-0.146}$~mb)
but it is still compatible with the data.  Our measurement does not
allow a bottom cross section to be deduced, which is predicted by
FONLL to be $\sigma_{b\bar{b}}^{FONLL} =
1.87^{+0.99}_{-0.67}$~$\mu$b.

In conclusion, we have measured single electrons from heavy flavor
decays in $p + p$ collisions at $\sqrt{s} = 200$~GeV.  These data
provide a crucial benchmark for pQCD heavy quark calculations.  We
observe that above $p_T \approx 2$~GeV/$c$ the electron spectrum is
significantly harder than predicted by a LO PYTHIA charm and bottom
calculation.  Contributions to the charm production cross section in
excess of the considered FONLL calculation, {\it e.g.} from jet
fragmentation, cannot be excluded.  Similar excess at high $p_T$ was
observed by Fermilab experiments (CDF and D{\O}) at $\sqrt{s} =
1.96$~TeV ~\cite{acosta03}.  The new data reported here provide an
important baseline for the study of possible medium modification of
heavy quark production at RHIC.


We thank the staff of the Collider-Accelerator and Physics
Departments at BNL for their vital contributions.  We acknowledge
support from the Department of Energy and NSF (U.S.A.),
MEXT and JSPS (Japan), CNPq and FAPESP (Brazil), NSFC (China),
CNRS-IN2P3 and CEA (France),
BMBF, DAAD, and AvH (Germany),
OTKA (Hungary), DAE and DST (India), ISF (Israel),
KRF and CHEP (Korea), RMIST, RAS, and RMAE (Russia),
VR and KAW (Sweden), U.S. CRDF for the FSU,
US-Hungarian NSF-OTKA-MTA, and US-Israel BSF.

\def\IJMPA{{Int. J. Mod. Phys.}~{\bf A}}
\def\JPG{{J. Phys}~{\bf G}}
\def\NCA{Nuovo Cimento}
\def\NIM{Nucl. Instrum. Methods}
\def\NIMA{{Nucl. Instrum. Methods}~{\bf A}}
\def\NPA{{Nucl. Phys.}~{\bf A}}
\def\NPB{{Nucl. Phys.}~{\bf B}}
\def\PLB{Phys. Lett. B}
\def\PLC{Phys. Repts.\ }
\def\PRL{Phys. Rev. Lett.\ }
\def\PRD{Phys. Rev. D}
\def\PRC{Phys. Rev. C}
\def\ZPC{{Z. Phys.}~{\bf C}}
\def\etal{{\it et al.} }


\begin{references}
\bibitem{pythia_tuned}
  We used PYTHIA 6.205 with a modified set of parameters~\cite{phenix_auau130}
  and CTEQ5L parton distribution functions~\cite{cteq5l}.
\end{references}
\end{document}